\documentclass[12pt]{JHEP3}
\usepackage{epsfig}
\preprint{ {\tt hep-th/0304227} \\ {IC/2003/28} }
\newcommand{\be}[1]{ \begin{equation}\label{#1} }
\newcommand{\ee}{\end{equation}}
\newcommand{\bea}[1]{\begin{eqnarray}\label{#1} }
\newcommand{\eea}{\end{eqnarray}}
\newcommand{\eq}[1]{(\ref{#1})}
\newcommand{\del}{\partial}
\title{ Konishi anomaly approach to gravitational $F$-terms}
\author{Justin R. David$^a$ , Edi Gava$^{a, b}$, K.S. Narain$^{a}$ \\
$^a$High Energy Section, \\
 The Abdus Salam International Centre for Theoretical Physics, 
\\Strada Costiera, 11-34014 Trieste, Italy.\\
$^b$Instituto Nazionale di Fisica Nucleare, sez. di Trieste, \\
and SISSA, Italy. \\
\email{justin, gava, narain@ictp.trieste.it}
}
\abstract{
We study gravitational corrections to the  effective superpotential in 
theories with a single adjoint chiral multiplet, using the generalized 
Konishi anomaly and the gravitationally
deformed chiral ring. We show that the genus one
correction to the loop equation in the corresponding matrix model agrees with
the gravitational corrected anomaly equations in the gauge theory. 
An important ingrediant in the proof is the lack of factorization of chiral
gauge invariant operators in presence of a supergravity background. We
also find a genus zero gravitational correction to the superpotential,
which can be removed by a field redefinition. }

\begin{document}

\baselineskip 3.5ex

\section{Introduction}
Recently much progress has been made in the computation 
and understanding of F-terms which describe the coupling
of $N=1$ gauge theories to $N=1$ supergravity 
\cite{Dijkgraaf:2002dh,Dijkgraaf:2002yn,Klemm:2002pa,
Ooguri:2003qp,Ooguri:2003tt}. 
In particular the
original conjecture of Djikgraaf and Vafa \cite{Dijkgraaf:2002dh}, 
relating these $F$-terms to non-planar corrections to the free energy of
a related matrix model has been proved by \cite{Ooguri:2003tt} 
using diagrammatic techniques, extending to the gravitational case
the super-Feynman diagram techniques of \cite{Dijkgraaf:2002xd}
Crucial ingredient in the proof
was the modification of the chiral ring relations due to
the coupling of the gauge theory to supergravity. In particular,
if one restricts to the first non-trivial gravitational $F$-term
contribution, corresponding to the genus one correction 
in the related matrix model, one needs to take into account just
the modification which follows from standard 
$N=1$ supergravity tensor calculus. 
The purpose of the present work is rather to study the problem   
from the viewpoint of generalized Konishi anomaly relations in the
chiral ring, by extending to the case of $N=1$ gauge theories 
coupled to $N=1$ supergravity the strategy of \cite{Cachazo:2002ry}
The key point in our analysis will be,together with
the modification of the chiral ring mentioned above, the
observation that in the presence of a non-trivial supergravity
background the usual factorization property of chiral correlators
does not hold. In particular connected two point functions
are generically non-vanishing, much like in matrix models, where     
connected correlators receive a subleading $1/N^2$ contribution, in
the $1/N$ expansion  

The paper is organized as follows: in section 2 we study 
in detail the deformation of the chiral ring due to the
coupling to supergravity and some of the resulting 
relations that we will need subsequently. In section 3
we estimate the chiral, connected two- and three-point
functions that will enter later in the equations obtained
using the generalized Konishi anomaly. Section 4 contains 
the main results: we derive there the generalized loop equations
which enables us to solve for the relevant connected two-point
functions. We show the uniqueness of the solution and use it
to compute the first non-trivial gravitational correction  
to the effective superpotential.. We finally show that 
this agrees with the genus one correction to the matrix
model free energy. 
In section 5 we make our conclusions and mention some
open problems.

\section{The chiral ring}

A basic ingredient in deriving the  effective $F$-terms in $N=1$  gauge
theories after integrating out the adjoint matter is the chiral ring 
\cite{Cachazo:2002ry}. 
All quantities in the chiral ring are defined modulo
$\bar{D}_{\dot{\alpha}}$ exact terms, here $\bar{D}_{\dot{\alpha}}$ 
refers to the differential operator conjugate to the supersymmetric
current
$\bar{Q}_{\dot{\alpha}}$. 
It is sufficient to restrict the quantities to be in the chiral
ring as $\bar{D}_{\dot{\alpha}}$ exact terms do not contribute to $F$
terms. For  the $N=1$ gauge theory with adjoint matter on $R^4$ the chiral
ring relations are  given by
\be{flat-chi}
[W_\alpha, \Phi] = 0 \;\; \hbox{mod } \bar{D}, \;\;\;\;\;
\{W_\alpha, W_\beta\} =0 \;\; \hbox{mod } \bar{D},
\ee
where  $W_\alpha$ is the $N=1$ gauge multiplet and $\Phi$ is the
chiral multiplet in the adjoint representation of the gauge
group.

The chiral ring relations given above are modified in the presence of
a curved background \cite{Ooguri:2003tt}.
We review here the derivation of the modified chiral ring in presence
of background $N=1$ gravity. 
Consider the following $\bar{D}$ exact quantity
\be{dering}
\{ \bar{D}^{\dot{\alpha}}, [D_{\alpha\dot{\alpha}}, W_\beta ]\} =
\{ [\bar{D}^{\dot{\alpha}}, D_{\alpha\dot{\alpha}}], W_\beta \}
\ee
where $D_{\alpha\dot{\alpha}}$ is the full covariant derivative
containing the gauge field and the spin connection. 
We used the Jacobi identity and the fact that $W_\beta$ is chiral
to obtaining the second term in the above equality. From the Bianchi
identity \cite{Wess:1992cp} for covariant derivatives one has
\be{bianchi}
[\bar{D}^{\dot{\alpha}}, D_{\alpha\dot{\alpha}}]  = 4i W_\alpha - 8i
G_{\alpha\beta\gamma} M^{\beta\gamma}
\ee
where $G_{\alpha\beta\gamma}$ is the $N=1$ Weyl multiplet and
$M^{\alpha\beta}$ refers to the Lorentz generator on spinors,
its action on a spinor is given by 
\be{lgen}
[M^{\alpha\beta}, \psi_\gamma]=
\frac{1}{2}(\delta^\alpha_\gamma \psi^\beta + \delta^\beta_\gamma
\psi^\alpha) 
\ee
Substituting  the Bianchi identity \eq{bianchi} in
\eq{dering} and using the action of the Lorentz generator 
we find the deformed chiral ring given below
\bea{ring}
\{W_\alpha, W_\beta \} = 2G_{\alpha\beta\gamma} W^\gamma \;\; \hbox{ mod }
\bar{D}, \;\;\;\;\;
\{W^\alpha, \Phi \} = 0  \;\; \hbox{ mod } \bar{D}.
\eea
The second equation is obtained by replacing the $W_\beta$ in
\eq{dering} by $\Phi$. For the conventional $N=1$ supergravity theory,
in the first equation above, only the
$SU(N)$ part of the gauge field $W^{\gamma}$ appears in the right hand
side. In \cite{Ooguri:2003tt}, the modification of the ring 
involved the $U(1)$ part of the gauge field as well, which corresponds
to a non-standard $N=1$ supergravity theory relevant to 
D-brane gauge theories. In the present paper we will always be
restricting ourselves to the standard $N=1$ supergravity. In order to
avoid explicitly including the $SU(N)$ projectors in all the formulae
below, we shall always take gauge field backgrounds to be in the
$SU(N)$ part of $U(N)$.

Using the deformed chiral ring we can derive many identities 
valid in the ring which
are used crucially  in the next sections \footnote{The
conventions followed in this paper are given in Appendix A.}.
From the definition of $W^2$ and  \eq{ring} we have
\bea{ident1}
W_{\alpha} W_\beta &=& \frac{\epsilon_{\alpha\beta}}{2} W^2 +
G_{\alpha\beta\gamma} W^{\gamma},  \\ \nonumber
W_\alpha W_\beta W_\gamma &=&
\frac{1}{2} \epsilon_{\alpha\beta} W^2 W_\gamma +
\frac{1}{2}G_{\alpha\beta\gamma} W^2 +
G_{\alpha\beta\delta}G^{\delta}_{\;\;\gamma \sigma} W^\sigma, 
\eea
using the above identities we are led to the following relations 
\bea{ident}
W_{\alpha} W^2 &=& -\frac{1}{2}   W^2W_\beta - 
\frac{1}{2} G^2 W_\alpha, \\ \nonumber
W^2 W_\alpha &=& 
-\frac{1}{2} W_\alpha W^2 - \frac{1}{2} G^2 W_\alpha, \\ \nonumber
[ W^2, W_\alpha] = 0, \;\; &\;&
 W^2  W_\alpha = -\frac{1}{3} G^2 W_\alpha, \;\;\;\;
W^2 W^2 = -\frac{1}{3} G^2 W^2 , \\ \nonumber
(G^2)^2 &=&
G^{\alpha\beta\gamma} G_{\alpha\delta\sigma}
G^{\sigma}_{\;\;\beta\rho} G^{\delta\rho}_{\;\;\;\;\gamma}.
\eea

These identities imply that the gauge invariant combination of
certain chiral operators vanish in the chiral ring.
The following  chiral operator vanishes in the chiral ring.
\be{op3}
G_{\alpha\beta\gamma} \rm{Tr}( W^\gamma\Phi \Phi \ldots) = 0. \;\;\;\;
\hbox{mod } \bar{D}
\ee
It is clear that if there are no $\Phi$'s in the trace, the above
equation is true for the gauge group $SU(N)$. To proof the above
identity for arbitrary number of $\Phi$ we use the following equation
\bea{proof3}
\rm{Tr} (W_\alpha W_\beta \Phi \Phi \ldots) &=& - \rm{Tr} 
(W_\beta W_\alpha \Phi\Phi \ldots) , \\ \nonumber
&=& \frac{1}{2} \epsilon_{\alpha\beta} \rm{Tr} (W^2 \Phi \Phi \dots).
\eea
To obtain the first equation above we have used the cyclic property of trace
and \eq{ring}. Now multiplying the first equation of \eq{ident1} with
arbitrary number of $\Phi$'s and using \eq{proof3} we obtain \eq{op3}.
Multiplying \eq{op3} by $G^{\alpha\beta}$ and using \eq{con2} we
obtain
\be{op1}
G^2 \rm{Tr}( W_\alpha \Phi \Phi \ldots ) = 0, \;\;\;\; \hbox{mod }
\bar{D}.
\ee
Another important identity in the chiral ring is 
\be{op2}
G^4 = (G^2)^2 =0, \;\;\;\; \hbox{mod  } \bar{D}
\ee
The proof goes along the same lines as  the derivation of the deformed
chiral ring. Consider the following $\bar{D}$ exact quantity
\be{dexact}
\{\bar{D}^{\dot{\alpha} }, [D_{\alpha\dot{\alpha} },  G_{\beta\gamma\delta}]
\} = \{[
\bar{D}^{\dot{\alpha} }, D_{\alpha\dot{\alpha}}],
G_{\beta\gamma\delta} \}
\ee
As $G_{\beta\gamma\delta}$ is uncharged with respect to gauge field
the covariant derivative $D_{\alpha,\dot{\alpha}}$ contains only the
spin connection. The Bianchi identity for covariant derivatives now implies
\be{bianc2}
[\bar{D}^{\dot{\alpha}}, D_{\alpha\dot{\alpha}} ] =
-8iG_{\alpha\beta\gamma}M^{\beta\gamma}
\ee
Substituting the above equation in \eq{dexact} we obtain the following
equation in the chiral ring
\be{symmg}
G_{\alpha\beta\sigma} G^{\sigma}_{\;\gamma\delta} +
G_{\alpha\gamma\sigma} G^{\sigma}_{\;\beta\delta} +
G_{\alpha\delta\sigma} G^{\sigma}_{\;\beta\gamma}  =0
\ee
Multiplying this equation by 
$G^{\alpha\gamma\rho} G^{\beta\delta}_{\;\;\;\rho}$ 
so that all the free indices are
contracted and using the last equation in \eq{ident} and \eq{con2} we
obtain \eq{op2}. As a result the gravitational corrections to the
$F$-terms
truncate at order $G^2$.


\section{Estimates of connected part of correlators}

In the absence of gravity correlators of gauge invariant operators
factorize in the chiral ring \cite{Cachazo:2002ry}. 
This factorization enabled one to show that the loop equations
satisfied by the resolvent on the gauge theory agreed with the 
loop equations of the matrix
model in the large $N$ limit. On the matrix model side factorization
in the loop equations was due to the the large $N$ limit.
However the correspondence of the gauge theory with the matrix model
proposed by  Dijkgraaf and Vafa in \cite{Dijkgraaf:2002dh} 
goes beyond the large $N$
approximation. If the $N=1$ gauge theory is placed in a background
gravitational field then the gravitational corrections to the $F$
terms of the gauge theory is of the form
\be{gravcor}
\Gamma_1 = \int d^4x d^2\theta G^2 F_1(S),
\ee
here $S$ refers to the gaugino condensate.
The Dijkgraaf and Vafa proposal states that the one can calculate $F_1$
from the genus one diagrams of the corresponding matrix model.
It has been shown in 
\cite{Ooguri:2003tt} that the modification
of the chiral ring in the presence of gravity 
allows the computation of the gravitational corrections to the
$F$-term. The gravitational corrections enter at genus one on the
gauge theory side 
and they reduce to the genus one diagrams of the corresponding
matrix model. This implies that the loop equations of the gauge theory
would not factorize in presence of gravity, as the loop equations of
the matrix model do not factorize at genus one. 
Therefore a priori we expect that
gauge invariant operators do not factorize in the presence of gravity.
In this section we estimate the gravitational corrections to the
connected parts of various correlators that can appear in the loop,
equations using the deformation of the chiral ring in presence of 
gravity \eq{ring}. 

To estimate the gravitational corrections to the $F$-term obtained by
integrating out the chiral multiplet we require the two point function
the chiral scalars $\langle \Phi \Phi \rangle$.
This is obtained 
after integrating out the antichiral scalar and  it is given by
\be{2pt}
\langle \Phi( x, \theta) \Phi(x', \theta') \rangle =
\frac{ \bar{m}}{ - \Box + \frac{i}{2} D^\alpha W_\alpha + 
i W^\alpha D_\alpha + m\bar{m} } 
\delta^4(x-x') \delta^2(\theta
-\theta')
\ee
This propagator in the presence of a gravitational background was
derived in \cite{Gates:1983nr} and in writing the above equation
we have assumed that the gravitational
background is on shell which allows one to set the other terms which occur
in the propagator to zero, 
$\Box$ is \eq{2pt} stands of the full gauge and
gravitational covariant derivative. The action of $W_\alpha$ is
dictated by the representation of $\Phi$, in this paper we will
restrict ourselves to the adjoint action. The delta functions in
\eq{2pt} refer to the full covariant delta function in curved
superspace.
In order to obtain the leading estimates for the connected component of various
correlators due to the modification of the chiral ring
it is sufficient to use the free d' Alembertian operator
and a constant gaugino background.
We argue this as follows, we can expand the propagator of \eq{2pt} 
in a weak field as
\bea{expweak}
\frac{\bar{m} }{-\Box + i D^\alpha W_\alpha + m \bar{m} }  &=&
\frac{\bar{m}}{-\Box_0 + iW^\alpha D_\alpha + m\bar{m} } \\ \nonumber
&+& 
\frac{\bar{m}}{-\Box_0 + iW^\alpha D_\alpha + m\bar{m} }  
( \Box - \Box_0 ) 
\frac{1 }{-\Box_0 + i W^\alpha D_\alpha + m\bar{m} }  
+ \ldots
\eea
here $\Box_0$ refers to the free d'Alembertian operator. We have also
dropped the terms $D^\alpha W_\alpha$
in \eq{2pt} as we have considered a covaraintly constant gaugino background.
From the expansion we see that the corrections in using
the free d' Alembertian operator in the propagator 
always occur with the factor $(\Box -\Box_0)$, 
which is proportional to
the gravitational background and therefore subleading to the estimate
obtained using the free d' Alembertian operator in the first term. One
has to make a similar expansion for the covariant delta function in 
\eq{2pt} and again one can see corrections in using the 
flat space delta function are subleading.
However we will see later that if  there is no gravitational
contribution form the deformed chiral ring  the leading estimate for
the connect part of correlators arise from 
corrections in the propagator due to the presence of the full
covariant $\Box$ and the covariant delta function.
For the free d'Alembertian operator in the propagator 
it is possible to go over to
momentum space and to write a Schwinger parameterization of the
propagator as follows.
\be{prop}
\langle \Phi(x, \theta)  \Phi (x', \theta ) \rangle
= \int   ds  d^4 p d^2 \pi  e^{ip(x-x')  }  e
^{-\frac{s}{\bar{m}} (p^2 + W^\alpha \pi_\alpha + m\bar{m})}
\ee
where $\pi^\alpha = iD^\alpha$. In the above equation
we have restricted to the superspace variable $\theta$ to be the
same at $x$ and $x'$ as we will be interested in correlators at the
same point in the superspace variable $\theta$.
We will now use this propagator and the modified chiral ring \eq{ring}
to make estimates for the connected part of various correlators.
The modified ring allows more than two insertions of $W_\alpha$ in a
given index loop, using the identities in \eq{ident} such
contributions can be converted to gravitational corrections.
At this point one might wonder if contributions to the connected diagrams
of gauge invariant operators in presence of a gravitational background 
are in contradiction with
the result found in \cite{Witten:1994ev}. There it was found that on an
arbitrary K\"{a}hler manifold gauge invariant operators of the $N=1$ theory
factorize. The background considered in \cite{Witten:1994ev} 
was entirely bosonic,
we find the estimates of 
contribution to the connected diagram indeed vanish for 
a purely bosonic background, thus there is no contradiction with
\cite{Witten:1994ev}. 
We will indicate this as we evaluate the estimates of various
correlators.

The various operators involved in the correlators of
interest are
\bea{defoper}
{\cal R}(z)_{ij} = - \frac{1}{32 \pi^2}
\left(\frac{W^2}{z-\Phi} \right)_{ij},  &\;\;\;& R(z) =
{\rm Tr}{\cal R}(z) 
\nonumber \\
 \rho_{\alpha~ij}(z)= \frac{1}{4\pi}
 \left(\frac{W_{\alpha}}{z-\Phi}\right)_{ij}, &\;\;\;&
w_{\alpha}(z)={\rm Tr} \rho_{\alpha}(z) \nonumber\\
{\cal T}(z)_{ij}= \left(\frac{1}{z-\Phi}\right)_{ij}, 
&\;\;\;&  T(z)={\rm Tr}{\cal T}(z)
\eea 
here we have defined separate symbols for the matrix elements and the
trace for later convenience, the gauge invariant operators we will
consider are the ones with the trace in the above equation.
The contour integrals of $R$, $w_{\alpha}$ and $T$ around $i$-th
branch cut define  the gaugino bilinear $S_i$, the $U(1)$ gauge field 
$w_{\alpha i}$ and $N_i$ respectively in $U(N_i)$ subgroup of $U(N)$
as in \cite{Cachazo:2002ry}. The fact that we are here restricting the
background gauge field to be in $SU(N)$ rather than $U(N)$ 
implies that $\sum_i w_{\alpha i} =0$. The chiral ring relation $G^2
w_{\alpha}(z)=0$ implies that $G^2 w_{\alpha i}=0$ for all $i$.  

We first consider estimates of the connected part of two point function,
we will discuss in detail the estimate for the follow correlator
\be{rr}
\langle R(z, x, \theta), R(w, y, \theta) \rangle_c, 
\ee
where the subscript stands for the connected part and out line the
derivation of the estimates for the other two point functions.
The various contribution to this correlator in \eq{rr} can be found by
expanding in $z$ and $w$, by definition of the connected correlator
the expansion starts of with the power $1/z^2w^2$. 
Let us focus on a
Feynman diagram consisting of $l$ loops, 
there will be $l+1$ bosonic and fermionic momentum
integrations in this diagram. 
The extra momentum integral comes from
the final Fourier transform to convert to the position space
representation for the  above correlator. 
We can organize this diagram 
into index loops due to the adjoint action of $W_\alpha$. 
The fermionic momentum integral forces us to bring down $2(l+1)$
powers of $W$ from the propagator in \eq{prop}. 
To obtain the leading gravitational correction we would like the
number of index loops to be as large as possible so that we can avoid
having more than two $W$'s in a given index loop.
The number
of index loops $h$ and the number of loops are related by $ l= h-1 +
2g$, where $g$ is the genus of the diagram. For a given number
of Feynman loops the number of index loops is largest for genus zero,
thus the leading estimate  to the connected graph arises from the planar
diagram.  For
a planar diagram  we need to saturate the fermionic momentum integrals by
bringing down $2h$  $W$'s. This can be done by inserting $W^2$ in $h$
index loops. We still have two more external $W^2$  in \eq{rr}. This
can at best be inserted in two different index loops. Thus we have two
index loops with $(W^2)^2$ insertions. Using the identities in
\eq{ident} we see that each of them reduces to $G^2 W^2$. Thus 
there is a term proportional to  $G^4 W^2$ on one of the index loops.
Note that if one had a purely bosonic background both $G$ and $W$
would start  at $\theta$ in the superspace expansion, thus $G^4W^2$
would vanish in agreement with \cite{Witten:1994ev}. In fact
$G^4W^2$, as is trivial in the chiral ring by \eq{op2},  
the leading estimate for
the correlator in \eq{rr} vanishes. In the next section it is
shown that the two point function in \eq{rr} in fact vanishes.
Next we consider the following two point function 
\be{rrho}
\langle R(z,x, \theta) w_\alpha(w, x', \theta) \rangle_c,
\ee
we can apply the same counting again, finally in the planar diagram we
will be left with at best with one index loop with $(W^2)^2$ and one
with $W^2 W_\alpha$ insertions. This reduces to a $G^2 W^2$ insertion
and a $G^2 W_\alpha$ insertions, which implies that the leading
gravitational contribution to \eq{rrho} is proportional to $G^4$.
Now we have seen in \eq{op1} that
$G^2 W_\alpha$ is zero in the chiral ring, thus this leading estimate
in fact vanishes in the chiral ring.
Similarly, consider the correlator
\be{rt}
\langle R(z, x, \theta) T(w, x', \theta) \rangle_c.
\ee
Here we will be left with $(W^2)^2$ in a single index loop, which
reduces to $G^2 W^2$. Thus the above correlator is proportional to
$G^2 W^2$. For a purely bosonic background we see that this contribution
again vanishes, consistently with \cite{Witten:1994ev}. For the case of
\be{rhorho}
\langle w^\alpha(z,x, \theta)  w_\alpha(w, x', \theta) \rangle_c, 
\ee 
we will be left with either $W^\alpha W^2$ insertion in two different
index loops or a $(W^2)^2$ insertion in a single index loop. The
former case vanishes in the chiral ring, but the latter case survives,
with a contribution proportional to $G^2 W^2$. 
For the following two point function
\be{rhot}
\langle w_\alpha (z, x, \theta) T(w, x', \theta) \rangle_c, 
\ee
there is a $W^\alpha W^2$ insertion in a single index loop, which is 
proportional to $G^2 W_\alpha$. Note that this leading contribution
vanishes in the chiral ring due to \eq{op1} and also for a purely
bosonic background.
Finally, we have the two point function
\be{tt}
\langle T(z, x, \theta) T(w, x', \theta) \rangle_c
\ee
For this case all the $h$ index loops are saturated with one $W^2$ and
there are extra insertions of $W^2$ for any of the $h$ index loops, 
as there
is no external $W$. Thus we have no contribution for the connected
part of this correlator from the modified ring. However we will see
later that there is a direct gravitational contribution to the above
correlator. This can be seen roughly  as follows: the d' Alembertian in
\eq{2pt} carries the covariant derivatives which can possibly
contribute to the connected two point function,  
as seen in the
expansion of the propagator in \eq{expweak}. 
This fact can be further justified by the evaluation of 
the 1-loop effective action obtained by integrating out the chiral multiplet 
in the absence of the gauge field background, 
which gives a term proportional
to $G^2\ln(m)$ \cite{Gates:1983nr}. 
Therefore we expect the leading term in the  correlator in 
\eq{tt} to be proportional to $G^2$ and  this will be shown 
explicitly in the next section. Again we see that for a purely bosonic
background $G^2$ is proportional to $\theta^2$, which implies that the
lowest components of the superfields in \eq{tt} factorize consistently
with \cite{Witten:1994ev} \footnote{Note that the fact that the lowest
component factorize cannot be used to promote it to a superfield
equation as the is no $Q_\alpha$ which preserves the background.}.
We summarize the estimates of the 
various connected two point correlators in the
following table for future reference

\begin{center}
\begin{tabular}{l l }
\\
\hline
Correlator & $\;\;\;$ Estimate
\\ \hline
$\langle R R\rangle_c$  & $G^4 =0$  \\
$\langle R w_\alpha\rangle_c$ 
& $G^4 =0$\\
$\langle R T\rangle_c$ & $G^2S^h$ \\
$ \langle w^\alpha w_\alpha \rangle_c$ & 
$G^2 S^h$  \\
$\langle w_\alpha T\rangle \rangle_c$  
& $G^2 S^{h-1} w_{\alpha~i}= 0$\\
$\langle T T \rangle_c$ & $G^2 S^{h-1}$ \\
\hline
\\
\end{tabular}\\
 Table 1. \\
\end{center}
\noindent
where $S$ represents schematically any of the $S_i$'s, and we have
used the chiral ring relations $G^4=0$ and $G^2 w_{\alpha i}=0$. 

Now we provide estimates for the fully connected part of various three point
functions. 
All the fully connected part of the three point
functions are proportional to at least $G^4$, therefore using \eq{op2}
they all vanish in the chiral ring. 
We discuss  the method of arriving at the estimates for one
case in detail and
just outline the results for the others. Consider the following fully
connected three point function.
\be{rrr}
\langle R(z, x, \theta) R(w, x', \theta) R(u, x'', \theta) \rangle_c.
\ee
By the definition of the full connected three point function, 
the first possibly non zero term of the 
expansion in $z, w, u$ starts at $1/(zwu)^2$. Consider a contribution
to any of the correlators appearing in this expansion.
A Feynman diagram consisting of $l$ loops will now have $l+2$ 
bosonic and fermionic momentum
integrations. This is because a three point function 
in momentum space will in general 
have two external independent momenta, and then converting that to
position space will involve these additional momentum integrals. As we
have argued earlier for the case of the two point function, the
leading contribution will be from the genus zero graphs.
For a planar graph then there are $2(h+1)$ fermionic momentum integrals
to be done. Therefore in addition to inserting $h$ index loops by
$W^2$, at best four different index loops will have $(W^2)^2$ insertions 
(we assume $l$ is large enough).
Using the identities in \eq{ident} we see that 
each $(W^2)^2$ insertion is proportional to $G^2 W^2$. Thus the above
three point function 
is proportional to $G^8$, but $G$ is fermionic and has $4$ independent
components, thus $G^8$ vanishes due to Fermi statistics.
Similar arguments show that the following correlators vanish
\bea{3ptvanish}
\langle R (z, x, \theta) R(w, x', \theta) w_\alpha(u, x'', \theta)
\rangle_c =0, \\ \nonumber
\langle R( z, x, \theta) R(w, x', \theta) T(u, x'', \theta) \rangle_c
=0, \\ \nonumber
\langle R( z, x, \theta) w^\alpha(w, x', \theta) w_\alpha(u, x'', \theta) 
\rangle_c =0, \\ \nonumber
\langle R( z, x, \theta) w^\alpha(w, x', \theta) T (u, x'', \theta) 
\rangle_c =0, \\ \nonumber
\eea
The first correlator in \eq{3ptvanish} is proportional to $G^8$ and the rest 
are  proportional to $G^6$, thus they vanish due to Fermi statistics.
Now consider the following three point function
\be{3ptvan2}
\langle w^\alpha( z, x, \theta) T(w, x', \theta) T (u, x'', \theta) 
\rangle_c . \\ \nonumber
\ee
We have seen that in a planar graph 
the fermionic momentum integrations force one to insert at least one factor 
of $W^2$ in all of the $h$ index loops and 
there is at least one index loop with a $(W^2)^2$ insertion. 
Using the identities in \eq{ident} this can be manipulated to a
gravitational contribution proportional to $G^2W^2$.
For the above correlator 
there is one more  index loop with an insertion of $W^2W^\alpha$ and  
again using
the identities in \eq{ident}, this term is proportional to $G^2
W^\alpha$. Thus the leading gravitational contribution to the three
point function in \eq{3ptvan2} is proportional to $G^4W_\alpha$ 
and thus it is zero in the chiral ring using \eq{op1}.
Next we consider the following three point function
\be{rtt}
\langle R( z, x, \theta) T(w, x', \theta) T (u, x'', \theta) 
\rangle_c , \\ \nonumber
\ee
As discussed above, since there are $2(h+1)$ momentum integration, all
the $h$ index loops have at least a $W^2$ insertion with one having a
$(W^2)^2$ insertions, the above correlator has an external $W^2$.
Therefore the leading gravitational contribution arises with 
two different index loops each with  a $(W^2)^2$ insertion  and 
using the identities
in \eq{ident} this can be manipulated in the chiral ring to give a
factor of $G^4$.  
The following three point function is also proportional
to $G^4$
\be{rhorhot}
\langle W^\alpha( z, x, \theta) W_\alpha(w, x', \theta) T (u, x'', \theta) 
\rangle_c , \\ \nonumber
\ee
Here again due to the momentum integrations there is already a factor
of $G^2$,  the two external $W$'s can be inserted in another index 
loop, but this loop already has an insertion of $W^2$, which gives rise to
a contribution proportional to $G^2$. Thus the three point function in
\eq{rhorhot} is proportional to $G^4$. Therefore the correlators in
\eq{rtt} and \eq{rhorhot} vanish in the chiral ring due to \eq{op2}.
Finally, let us consider the following three point function
\be{ttt}
\langle T( z, x, \theta) T(w, x', \theta) T (u, x'', \theta) 
\rangle_c . \\ \nonumber
\ee
As in the case for \eq{tt} we can not estimate the $G$ dependence of
this correlator solely using the chiral ring. 
But from the fact that the correlator in \eq{tt} is
proportional to $G^2$ and since there is at least one index loop with
$(W^2)^2$ insertion, we can arrive at the conclusion that 
the above three point function will be
proportional to $G^4$. To sum up we have examined all 
possibly non-vanishing full connected three point
fully and found them to be  least proportional to $G^4$, 
and therefore using \eq{op2} they vanish in the chiral ring.

\section{Anomaly equations and matrix model loop equations}

In this section we will use the generalized Konishi anomaly to extract
the gravitational corrections to the effective superpotential. We
recall that the anomaly equations obtained in
\cite{Cachazo:2002ry} in the absence of gravitational fields are as follows
\begin{eqnarray}
\langle R(z) R(z) \rangle  &-& \langle {\rm tr}(V'(\Phi){\cal R}(z)) \rangle
= 0 \nonumber\\
2\langle R(z) w_{\alpha}(z) \rangle  &- &\langle {\rm tr}(V'(\Phi)
  \rho_{\alpha}(z) \rangle =0 \nonumber \\
2\langle R(z) T(z) \rangle  &-&\langle {\rm tr}(V'(\Phi) {\cal T}(z) \rangle  
+\langle w^{\alpha}(z) w_
{\alpha}(z) \rangle  =0
\label{Can}
\end{eqnarray}
Here $V$ denotes the classical superpotential of degree $n+1$.  We
have  indicated above the full two point functions that include the 
disconnected and connected two point functions. The latter vanish in
the absence of 
gravitational field but  as we will show below do not vanish in the 
presence of the gravitational field. These equations were obtained by 
the generalized Konishi anomalies upon transforming the adjoint chiral 
field  $\Phi$ as
$\delta \Phi_{ij}$ equal to ${\cal R}_{ij}(z)$,
$\eta^{\alpha}\rho_
{\alpha~ij}(z)$ and ${\cal T}_{ij}(z)$ respectively,  with
$\eta^{\alpha}$ 
being an arbitrary field independent spinor. In general for the 
infinitesimal transformation
$\delta \Phi_{ij} = f_{ij}$, the generalized Konishi anomaly is given by
\begin{equation}
\frac{\delta f_{ji}}{\delta \Phi_{k \ell}} A_{ij,k \ell}
\end{equation}
In the absence of gravitation 
\begin{equation}
A_{ij,k \ell} = (W^2)_{kj} \delta_{i\ell} + \delta_{kj} (W^2)_{i\ell} - 2
W^{\alpha}_{kj} W_{\alpha i\ell}
\label{an}
\end{equation}
Using the above anomaly and the equation $\{W_{\alpha} W_{\beta} \}
=0$ in 
the chiral ring, one obtains the equation 
(\ref{Can}).

In the presence of the gravitational field $G_{\alpha \beta \gamma}$, 
these equations are modified for two reasons:
firstly there is a direct gravitational anomaly (ie. even in the
absence of gauge fields, in
other words when chiral multiplets couple only to gravitational
fields) and secondly due to the modification of the ring \eq{ring}.

Under the infinitesimal transformation $\delta \Phi_{ij}= f_{ij}$ the 
pure gravitational contribution to the anomaly is
\begin{equation}
\frac{\delta f_{ji}}{\delta \Phi_{k \ell}} \alpha G^2 \delta_{kj} \delta_
{i\ell}
\end{equation}
where $\alpha =\frac{1}{3}$ as shown in the appendix B, and from now
on 
$G^2$ will denote $\;\;$ 
$\frac{1}{32\pi^2} G^{\alpha \beta \gamma} G_{\alpha \beta \gamma}$. This changes 
\begin{equation}
A_{ij,k\ell} \rightarrow A_{ij, k\ell} + \frac{1}{3} G^2 \delta_{kj} \delta_
{i\ell}
\label{Gan}
\end{equation}

It is easy to see that the pure gravitational anomaly and the
modification of 
the chiral ring (\ref{ring})  together with the consequent identities 
given in the equations \eq{ident},\eq{op1},\eq{op2},  give rise to the following modification of
the equations (\ref{Can}):
\begin{eqnarray}
\langle R(z) R(z) \rangle  &-&\langle tr( V'(\Phi){\cal R}(z)) \rangle   = 0 
\nonumber\\
2\langle R(z) w_{\alpha}(z) \rangle  &-&\langle tr(V'(\Phi) W_{\alpha}
(z)) \rangle   =0
\nonumber \\
2\langle R(z) T(z) \rangle  &-&\langle tr(V'(\Phi){\cal T}(z)) \rangle  
\nonumber\\
&+&\langle w^{\alpha}(z) w_
{\alpha}(z) \rangle   
 - \frac{1}{3} G^2 \langle T(z) T(z) \rangle  =0
\label{CanG}
\end{eqnarray}
Note that in the first two equations above the pure gravitational
anomaly cancels with the contributions coming from the
modification of the chiral ring via eq(\ref{ident}). This happens 
due to the remarkable 
fact that the pure gravitational anomaly (\ref{Gan}) comes with a 
factor of $\frac{1}{3}$ !! We have also used the fact that $G^2 w_\alpha$ 
vanishes in the chiral ring. In the last equation however there is no 
contribution due to the modification 
of the ring and  hence the last two terms on the left hand side arise 
solely from the pure gravitational anomaly.
Here again, a priori, the two point functions are the sum of connected
and disconnected parts.

For later purposes let us write the first equation in (\ref{CanG})
more explicitly
\begin{equation}
\langle R(z) \rangle^2 - V'(z)\langle R(z) \rangle - \frac{1}{4}f(z)
= -\langle R(z) R(z) \rangle_c 
\label{Req}
\end{equation}
where $f(z)$ is a polynomial of degree $n-1$ defined by
\begin{equation}
\langle {\rm tr} (V'(\Phi)-V'(z){\cal R}(z) \rangle =\frac{1}{4}f(z)
\end{equation}
and the subscript $c$ denotes the connected part of the correlation
function. The latter, as we have seen in the last section, goes as
$G^4$ and
therefore trivial in the chiral ring. As a result the equation for $R$
is unmodified by the gravitational field. Since the finite polynomial $f$
is determined completely by the periods of $R$, i.e. the contour
integrals around the various branch cuts $C_i$, $i=1,...,n$,
\begin{equation}
\frac{1}{2\pi i}\int_{C_i} dz R(z) = S_i
\end{equation}
we conclude that $R$ does not receive any gravitational corrections.

The strategy now is to expand all of the quantities appearing above
in a perturbation series in $G^2$. Of course, this series ends at
order $G^2$ since $G^4$ is trivial. Thus for example we 
write 
\begin{equation}
\langle T \rangle = T^{(0)} + G^2 T^{(1)} 
\end{equation}
and similarly for the connected parts of the 2-point functions
appearing in the above equations. As discussed in the last section,
the  latter start from order $G^2$, and therefore the equations for  $T^{(0)}$
and $w^{(0)}_{\alpha}$ are the same as in \cite{Cachazo:2002ry}. To go beyond this we
need to solve for the connected two point functions.

\subsection{Equations for the connected two point functions in the
  presence of 
gravitational fields}

We will now derive equations for the connected two point functions
that appear in eq.(\ref{CanG}). Although, in eq.(\ref{CanG}) the connected
2-point functions are of the form $\langle R(z) T(z)\rangle$, i.e. both
the operators are at same $z$, it turns out to be more convenient
to consider the two operators at different points in the complex
plane (say $z$ and $w$). The reason is that we can impose conditions
on a connected 2-point function, like $\langle R(z) T(w)\rangle$ that 
their integrals around various branch cuts in $z$ and $w$ plane 
vanish separately. As a result, we will be able to solve completely
the corresponding generalized Konishi anomaly equations.
              
We illustrate the general method of obtaining the generalized Konishi
anomaly equations for the connected 2-point functions in one example 
and then give the
complete set of equations which can easily be derived following the
methods given below.

Consider the infinitesimal transformation (local in superspace
coordinates $\;\;$ $(x^{\mu}, \theta, \bar{\theta})$)  
\begin{equation}
\delta \Phi_{ij} = {\cal R}_{ij}(z)  T(w)
\end{equation}
The Jacobian of this transformation has two parts
\begin{equation}
\frac{\delta (\delta \Phi_{ji})}{\delta \Phi_{kl}}= \frac{\delta
R_{ji}(z)}
{\delta \Phi_{kl}} T(w) + \sum_m R_{ji}(z) T_{mk}(w) T_{lm}(w)
\end{equation}
The first term in the equation above together with the variation of
the classical superpotential  gives rise to 
\begin{eqnarray}
\langle (R(z) R(z)& -& tr(V'(\Phi) {\cal R}(z) ) ) T(w)  \rangle  
\nonumber\\ &~&=  
\langle  (R(z) R(z) - tr(V'(\Phi){\cal  
R}(z) ) )  \rangle \langle T(w)  \rangle + 2\langle R(z) \rangle 
\langle R(z) T(w) \rangle _c \nonumber\\
&~&~-\langle   tr(V'(\Phi) {\cal R}(z) ) ) T(w) \rangle _c +
\langle R(z) R(z) T(w) \rangle _c
\label{c1}
\end{eqnarray}
where the subscript $c$ denotes completely connected 2 or 3 point
functions as indicated.  The first term on the right hand side
vanishes by virtue of the first equation of  (\ref{CanG}).

The second term in the Jacobian when combined with the anomaly 
(\ref{an},\ref{Gan}) gives rise to a single trace contribution 
\begin{equation}
\label{c2}
-\frac{1}{3} G^2 \langle tr({\cal R}(z) {\cal T}(w) {\cal T}(w)) \rangle  
=- \frac{1}{3}
G^2 \partial_w \frac{\langle R(z)\rangle -\langle R(w)\rangle}{z-w}
\end{equation}

Combining eqs.(\ref{c1}),(\ref{c2}) and the first equation of (\ref{CanG}), 
one obtains the following equation for the 
connected correlation functions:
\begin{equation}
(2 \langle R(z)\rangle - I(z))\langle R(z) T(w)\rangle_c + 
\langle R(z) R(z) T(w)\rangle_c-  \frac{1}{3}
G^2 \partial_w \frac{\langle R(z)\rangle -\langle R(w)\rangle}{z-w}=0
\label{rtc}
\end{equation}
Here the integral operator $I(z)$ denotes the following:
\begin{equation}
I(z) A(z)= \frac{1}{2\pi i}\int_{C_z} dy \frac{V'(y) A(y)}{y-z} 
\end{equation}
with the contour $C_z$ encircling $z$ and $\infty$. It is clear 
that for 
$A$ equal to ${\cal R}$, $\rho_{\alpha}$ or ${\cal T}$, the integral
operator reduces to 
\begin{equation}
I(z) A(z) = V'(\Phi) A(z)
\end{equation}
as is familiar in the matrix model works. 

Since the last term in the eq.(\ref{rtc}) is of order $G^2$ and 
involves one point function
of $R$ which is certainly not zero, we conclude
that the sum of the remaining terms which involve connected
correlations functions cannot all vanish
at order $G^2$. This proves our basic assertion. In fact the connected
3-pt. function $\langle R(z) R(z) T(w)\rangle_c$ vanishes, as argued in
eq.(\ref{3ptvanish}) in the last section, so eq.(\ref{rtc}) implies
that the connected 2-pt function $\langle R(z) T(w)\rangle_c$ does not
vanish at order $G^2$.

In order to completely solve the relevant connected correlation
functions, we need to consider all transformations of the form 
\begin{equation}
\delta \Phi_{ij} = A_{ij}(z) B(w)
\end{equation}
where $A$ is ${\cal R}$, $\rho_{\alpha}$ or ${\cal T}$ and $B$ is
$R$, $w_{\beta}$ or $T$. The resulting
generalized Konishi anomaly equations can be derived in the same way
as above and can be summarized in the following 
matrix equation:

\begin{eqnarray}  
\left[   \begin{array}{ccc}
{\hat M}(z)  &
2\langle T(z) \rangle
 & 0 \\
0 & M(z) &0 \\
0 & 0 & M(z)                       \end{array} \right]
      &~&  \left[
\begin{array}{ccc}
\langle T(z)T(w)\rangle_c &\langle T(z)R(w)\rangle_c &\langle T(z)
w_{\beta}(w)\rangle_c \\
\langle R(z)T(w)\rangle_c &\langle R(z)R(w)\rangle_c &\langle R(z)
w_{\beta}(w)\rangle_c \\
\langle w_{\alpha}(z)T(w)\rangle_c &\langle w_{\alpha}(z)R(w)\rangle_c 
&\langle w_{\alpha}(z)
w_{\beta}(w)\rangle_c \end{array}\right]
= \nonumber\\ &~& =
\frac{1}{3}G^2 \partial_w
\left[
\begin{array}{ccc}
T(z,w) & R(z,w) & 0\\
R(z,w) & 0 & 0 \\
0 & 0 & 5\epsilon_{\alpha \beta} R(z,w) \end{array}\right]
\label{matrix}
\end{eqnarray}

Here we have used the chiral ring equations $G^2 w_{\alpha} =0$ and 
$(G^2)^2 =0$ as shown in the section 2. We have also dropped various 
connected 3-pt functions that vanish via eq.(\ref{3ptvanish}). 
$M(z)$ denotes the integral
operator $(2\langle R(z)\rangle - I(z))$, ${\hat M}(z)$ denotes
$M(z)-\frac{2}{3}G^2 \langle T(z) \rangle$  and 
finally $R(z,w)$ and
$T(z,w)$ denote $(\langle R(z)\rangle - \langle R(w) \rangle)/(z-w)$ and
$(\langle T(z)\rangle - \langle T(w) \rangle)/(z-w)$ respectively.

There are a few points to note about these equations: \\
\noindent 1)  Chiral ring equations are consistent with the above
matrix equation. For example, if one takes the 
equation for 
$\langle w_{\alpha}(z) B(w)\rangle_c$ and multiplies by $G^2$ one
finds that the equation identically vanishes in the chiral ring.  \\
\noindent 2) We have used the estimates given in the last section only
to drop all the completely connected 3-pt. functions as they were
shown to go as $G^4$. The estimates for the connected 2-pt. functions
given in the Table 1, says that $\langle R(z)
R(w) \rangle_c$,  $\langle R(z) w_{\alpha}(w) \rangle_c$, and  $\langle T(z)
w_{\alpha}(w) \rangle_c$ all vanish in the chiral ring. We have not
used these estimates in the above equation but we note that the (1,3),
(2,2), (2,3), (3,1) and (3,2) matrix elements on the right hand side
vanish.
Thus the estimates given in the table are indeed consistent with the
matrix equation above. In fact, in the next subsection, we will show
that the solutions to this equation are unique thereby proving that
these connected 2-pt. functions vanish. Similarly had we used the
estimate for the 2-pt. function  $\langle T(z)T(w) \rangle_c$ which
was shown in the last section to go as $G^2$, we could have replaced
$\hat M$ by $M$ in the above equation.  \\  
\noindent 3) The integrability condition is satisfied: the above
equation is of the form
\begin{equation}
{\cal M}(z) N(z,w) =\partial_w  K(z,w)
\end{equation}
where ${\cal M}(z)$ is the first matrix operator appearing on the left hand
side of eq(\ref{matrix}), $N(z,w)$ and $K(z,w)$ satisfy $N(z,w)=
N^t(w,z)$ and $K(z,w) = K^t(w,z)$. The non-trivial consistency
condition then is
\begin{equation}
(\partial_w K(z,w)) {\cal M}^t(w) = {\cal M}(z) \partial_z K(z,w)
\label{int}
\end{equation}
The crucial identity needed for this is the one involving the integral 
operator $I(z)$ and is as follows:
\begin{equation}
(I(z) \partial_z-I(w) \partial_w)\frac{A(z)-A(w)}{z-w}  = (\partial_z-
\partial_w)\frac{I(z)A(z)-I(w)A(w)}{z-w} 
\end{equation}
for any function $A$ which is smooth at $z$ and $w$. This can be proved
by using the definition of the contours involved in $I(z)$ and $I(w)$. 
It follows that the (1,2), (2,1) and (3,3) components of the
integrability condition (\ref{int}) implies the following
equation:
\begin{equation}
G^2(\langle R(z) \rangle^2 - I(z)\langle R(z) \rangle )=0
\end{equation}
This equation is just $G^2$ times the first equation of (\ref{CanG}) if
one takes into account the fact that the connected part of the correlation 
function appearing in the latter already is of order $(G^2)^2$. The only
other non-trivial part of the integrability condition is its (1,1) component: 
\begin{equation}
G^2[(2\langle R(z) \rangle - I(z)) \langle T(z) \rangle
-\frac{1}{3}G^2  \langle T(z) \rangle^2] =0
\label{inttt}
\end{equation}
which is just $G^2$ times the disconnected part of the third equation of
(\ref{CanG}) thereby proving the integrability condition for the
(1,1) component. This is because all the connected parts appearing in that
equation will be trivial when multiplied by $G^2$. Note also that the
last term in eq.(\ref{inttt}) could have been dropped as it is
trivial.
Its origin comes from the extra term in $\hat M$ appearing in the
(1,1) component
of ${\cal M}$ which as argued in the point 2) above could have been
dropped. 

\subsection{Uniqueness of the solutions for the connected two point functions}

Since the integrability conditions are satisfied, solution to eq.(\ref{matrix})
exists. However the solution has a finite ambiguity which will be fixed
by the physical requirement that the contour integrals around all the branch
cuts of the connected two point functions in the the $z$ and $w$ planes
must vanish separately. The reason
for this is that the following operator equations hold: 
\begin{equation}
\frac{1}{2\pi i}\int_{C_i}dz R(z) = S_i, ~~~~ \frac{1}{2\pi i}
\int_{C_i} dw T(w) = N_i, ~~~~\frac{1}{2\pi i}\int_{C_i}dz w_{\alpha}(z) =
w_{\alpha~i}.
\end{equation}
where $S_i$ is the chiral superfield whose lowest component is the
gaugino bilinear in the $i$-th gauge group factor in the broken phase
$U(N) \rightarrow  \prod_{i=1}^n U(N_i)$ and 
$w_{\alpha~i}$ is the $U(1)$ chiral gauge superfield of the $U(N_i)$
subgroup. Since these fields are background fields, in the connected
correlation functions the contour integrals around the branch cuts
must vanish. Similarly since these background fields are independent
of the gravitational fields, order $G^2$ corrections to 
the one point functions of $R$, $w_{\alpha}$ and $T$ must also have
vanishing
contour integrals around the branch cuts.

For later use, we can write a complete set of normalized 
differentials using eq.
(\ref{Req}) upto order $G^2$  as
\begin{equation}
\omega_j = \frac{1}{4} \frac {dz}{[2\langle R(z)\rangle-V'(z)]}
\frac{\partial}
{\partial S_j}f(z),  ~~~~~~~~~~ \frac{1}{2\pi i}
\int_{C_i} \omega_j = \delta_{ij}
\label{basis}
\end{equation}   

To illustrate the method, we will again focus on
$\langle R(z) T(w) \rangle_c$. The action of operator $I(z)$ is given as:
\begin{equation}
I(z) \langle R(z) T(w) \rangle_c= V'(z) \langle R(z) T(w) \rangle_c +
\sum_{k=0}^{n-1} c_k(w) z^k
\end{equation}
where the first term on the right hand side comes from the contour integral 
around $z$ and the second term from that around $\infty$, with $n$ being the
order of $V'(z)$. Here we have used the fact that $R(z)$ asymptotically 
vanishes as  $1/z$ \footnote{ Actually the coefficient of $1/z$ is 
$\rm{Tr} W^2$ and hence in the connected 2-pt function it vanishes. 
As a result
the connected 2-pt function goes as $1/z^2$ (and for similar reasons $1/w^2$
asymptotically in $w$) which means that the sum over $k$ is between $0$ and 
$n-2$. However in the above expression we have kept the sum upto $n-1$ since,
as it will turn out, the condition of vanishing contour integrals around all 
the branch cuts will in particular imply that $c_{n-1}=0$}. The coefficients
$c_k(w)$ are arbitrary functions of $w$ which asymptotically vanish as $1/w^2$.
Similarly we have
\begin{equation}
I(w) \langle R(z) T(w) \rangle_c = V'(w) \langle R(z) T(w) \rangle_c +
\sum_{k=0}^{n-1} {\tilde c}_k(z) w^k
\end{equation}
with ${\tilde c}_k (z)$ being arbitrary functions of $z$ that vanish 
asymptotically as $1/z^2$. 

From (\ref{matrix}), the two equations that this two point function
satisfies are as follows:
\begin{eqnarray}
(2 \langle R(z)\rangle -I(z)) \langle R(z) T(w) \rangle_c&=&\frac{1}{3}G^2
\partial_w R(z,w) \nonumber\\
(2 \langle R(w)\rangle - I(w)) \langle R(z) T(w) \rangle_c&=&\frac{1}{3}G^2
\partial_z R(z,w) 
\label{rteq}
\end{eqnarray}
The solutions to these two equations are
\begin{eqnarray}
\langle R(z) T(w) \rangle_c &=& \frac{1}{2\langle R(z)\rangle -V'(z)}  
[\frac{1}{3}G^2 \partial_w R(z,w)+\sum_{k=0}^{n-1} c_k(w) z^k ] \nonumber\\
 &=& \frac{1}{2\langle R(w)\rangle -V'(w)}  
[\frac{1}{3}G^2\partial_z R(z,w)+\sum_{k=0}^{n-1} {\tilde c}_k(z) w^k ]
\label{rtcs}
\end{eqnarray}
Equating the two right hand sides , we see that $c_k$ and
$\tilde{c}_k$ are not arbitrary functions of the respective arguments
but are fixed upto a finite polynomial ambiguity in $z$ as well as
$w$. They must be of the form
\begin{eqnarray}
\sum_{k=0}^{n-1} c_k(w) z^k&=&\frac{1}{(2 \langle R(w)\rangle - 
V'(w))}\frac{G^2}{3}[\langle R(w)\rangle \frac{(V'(w)-V'(z)+(z-w)V''(w))}
{(z-w)^2} \nonumber\\
&~& +\frac{1}{4}\frac{(f(w)-f(z)+(z-w)f'(w))}{(z-w)^2} +
\sum_{k,\ell=0}^{n-1}c_{k\ell}z^k w^{\ell}]\nonumber\\
{\tilde c}_k(z) &=& c_k(z) + \sum_{\ell}(c_{k\ell}-c_{\ell k}) z^{\ell} 
\label{ck}
\end{eqnarray}
where $c_{k\ell}$ are arbitrary coefficients to be determined later.
In deriving the above equation we have used the equation (\ref{Req})
with the right hand side set to zero. This is because the 
correction coming from the right hand side is of order $G^4$ and hence
trivial.

Substituting this expression in eq.(\ref{rtcs}) and repeatedly using 
eq.(\ref{Req}), we obtain after some algebra:
\begin{eqnarray}
\langle R(z) T(w) \rangle_c &=& \frac{G^2}{3}\frac{1}{[2\langle R(z)
\rangle -V'(z)][2\langle R(w)
\rangle -V'(w)] }[\sum_{k,\ell=0}^{n-1}c_{k\ell}z^k
w^{\ell}+ \nonumber\\ &~&
\frac{[\langle R(z)\rangle \langle R(w)\rangle -V'(z)
\langle R(w)\rangle -\frac{1}{4} f(z)+ (z\leftrightarrow w)]}{(z-w)^2}]
\label{rtsolution}
\end{eqnarray}
Note that the second term in the bracket is symmetric in $z\leftrightarrow
w$.

As mentioned earlier, the connected two point function must obey the following 
conditions:
\begin{equation} 
\int_{C_i}dz\langle R(z) T(w) \rangle_c =\int_{C_i}dw
\langle R(z) T(w) \rangle_c =0
\label{contour}
\end{equation}
for all $i$, where $C_i$ is the contour around the $i$-th branch cut.

Let us first consider contour integrals around the branch cuts in $w$-plane.
To this end we can use the first equation in (\ref{rtcs}). The first term
on the right hand side is a total derivative in $w$ and therefore its contribution to the contour
integral vanishes. Thus we arrive at the condition:
\begin{equation}
\int_{C_i} c_k(w) =0
\label{intck}
\end{equation}
Using the expression (\ref{ck}) for $c_k$, and the fact that 
$w^{\ell}/(2\langle R(w)\rangle+V'(w))$ for $\ell=0,...,n-1$ form a complete
basis of holomorphic 1-forms in the present case, these equations  
determine $c_{k\ell}$ completely. Very explicitly if
\begin{eqnarray}
\sum_{k=0}^{n-1} t^{(i)}_k z^k &=& \int_{C_i}dw\frac{1}{(2 \langle R(w)
\rangle -
V'(w))}[\langle R(w)\rangle \frac{(V'(w)-V'(z)+(z-w)V''(w))}
{(z-w)^2} \nonumber\\
&~&~~~~~~~~~ +\frac{1}{4}\frac{(f(w)-f(z)+(z-w)f'(w))}{(z-w)^2}]
\label{tik}
\end{eqnarray}
then using the basis of normalized differentials eq.(\ref{basis}),
\begin{equation}
\sum_{k,\ell=0}^{n-1} c_{k\ell}z^k w^{\ell}= -\frac{1}{4}
\sum_{i=1}^n \sum_{k=0}^{n-1} 
t^{(i)}_k z^k \frac{\partial}{\partial S_i} f(w)
\label{cklsolution}
\end{equation}

We will now show that $c_{k\ell}$ are symmetric in $k$ and $\ell$ exchange.
Let us define a matrix $G_{ij}$ by the following equation 
\begin{equation}
G_{ij}=\frac{1}{2\pi i}\int_{C_j} dz \frac{1}
{2 \langle R(z)\rangle -V'(z)}\sum_{k=0}^{n-1}t^{(i)}_k z^k
\label{gij}
\end{equation}
We first simplify eq.(\ref{tik}) by using (\ref{Req}) as
\begin{equation}
\sum_{k=0}^{n-1} t^{(i)}_k z^k =- \int_{C_i}dw\frac{1}{(2 \langle R(w)
\rangle - 
V'(w))} \frac{2V'(z)V'(w)+f(z)+f(w)}
{4(z-w)^2} 
\end{equation}
where we have omitted a total derivative term with respect to $w$
since it does not contribute to the contour integral. Substituting
this in eq.(\ref{gij}) and noting that the residue of the first order pole 
$1/(z-w)$ vanishes due to eq.(\ref{Req}), we find that $G_{ij}$ is
symmetric. Eq.(\ref{gij}) can be solved explicitly
for $t^{(i)}_k$ as
\begin{equation}
\sum_{k=0}^{n-1}t^{(i)}_k z^k =\frac{1}{4} \sum_{j=1}^{n} G_{ij}
\frac{\partial}{\partial
S_j} f(z)
\end{equation}
Plugging this equation in eq.(\ref{cklsolution}) and using the fact
that $G_{ij}$ is symmetric, we find that the coefficients 
$c_{k\ell}$ are symmetric in $k$ and $\ell$ exchange as claimed
above. .

Finally note that the symmetry of $c_{k\ell}$ implies that $\langle R(z)
T(w)\rangle_c$ is symmetric in $z$ and $w$.This will be crucial in the
following. In particular this also implies that the contour integrals
around the branch cuts in the $z$-plane vanish.

To summarize this subsection, although we have discussed in detail the example
of $\langle R(z) T(w)\rangle_c$, it is easily seen that
the eq.(\ref{matrix}) and the conditions like eq.(\ref{contour}) 
fix the solutions for all the connected two point functions uniquely.

\subsection{Solutions for the connected two  functions}

Note that the right hand side of  equation (\ref{matrix}) vanishes for all
components except  $(1,1), 
(1,2),(2,1)$ and (3,3) (the last being proportional
to $\epsilon_{\alpha \beta}$). The uniqueness of the solution then implies
\begin{equation}
\langle R(z) R(w)\rangle_c=\langle w_{\alpha}(z) T(w)\rangle_c=
\langle w_{\alpha}(z) R(w)\rangle_c=\langle w_{(\alpha}(z) 
w_{\beta)}(w)\rangle_c=0
\end{equation}
This is in accordance with the estimates given in the Table 1.
We have already obtained the solution for $\langle R(z) T(w)\rangle_c$
in equations (\ref{rtsolution},\ref{tik},\ref{cklsolution}). Let us denote this
solution as $G^2 H(z,w)$. 

The remaining two equations are:
\begin{eqnarray}
M(z)\langle w^{\alpha}(z) w_{\alpha}(w)\rangle_c&=&\frac{10}{3}
G^2 \partial_w R(z,w) \nonumber\\
M(z)\langle T(z) T(w)\rangle_c &+& 2\langle T(z)\rangle 
\langle R(z) T(w)\rangle_c = \frac{1}{3}
G^2 \partial_w T(z,w)
\label{wandteq}
\end{eqnarray}
Comparing the first equation with that of $\langle R(z) T(w)\rangle_c$
namely eq.(\ref{rteq}) we conclude that
\begin{equation}
\langle w^{\alpha}(z) w_{\alpha}(w)\rangle_c=10 G^2 H(z,w)
\label{wwsolution}
\end{equation}
Finally taking the derivative of eq.(\ref{rteq}) with respect to 
$N_i \frac{\partial}{\partial S_i}$, and using the fact that to the
leading order (i.e. order(1)) $\langle T(z)\rangle =
(N_i \frac{\partial}{\partial S_i}+\frac{1}{2}w^{\alpha}_i
w_{\alpha~j}\frac{\partial^2}{\partial S_i \partial S_j})\langle
R(z)\rangle$, we
obtain the following equation:
\begin{equation} 
M(z)N_i \frac{\partial}{\partial S_i}\langle R(z) T(w)\rangle_c + 
2\langle T(z)\rangle 
\langle R(z) T(w)\rangle_c = \frac{1}{3}
G^2 \partial_w T(z,w)   
\end{equation}
where we have used the fact that $G^2 w_{\alpha~i}$ is trivial.
This equation is the same as the second equation of (\ref{wandteq}).
Uniqueness of the solution then implies that upto order $G^2$,
\begin{equation}
\langle T(z) T(w)\rangle_c =G^2 N_i \frac{\partial}{\partial S_i}H(z,w)
\end{equation}

We are now in a position to compute the gravitational corrections to the
one point functions of $R(z)$, $w_\alpha(z)$ and $T(z)$ from eq.(\ref{CanG}).
Note that in this equation the two point functions contain both the 
disconnected and connected pieces. Since $\langle R(z) 
R(z)\rangle_c$ and $\langle R(z) w_{\alpha}(z)\rangle_c$ vanish, 
the first two equations do not contain any connected parts. 
Uniqueness then implies that one point function of $R$ and $w_\alpha=
w_{\alpha}^i \frac{\partial}{\partial S_i}R$
do not get any gravitational correction. The non-trivial equation is the third
one. Using the results of this subsection we get
\begin{equation}
(M(z)-\frac{1}{3}G^2\langle T(z)\rangle)\langle T(z)\rangle +12 G^2 H(z,z) =0
\end{equation}
Expanding $\langle T(z)\rangle= T^{(0)}+ G^2 T^{(1)}$,
with $T^{(0)}= N_i \frac{\partial}{\partial S_i} R +\frac{1}{2}
w^{\alpha}_i w_{\alpha j}
\frac{\partial}{\partial S_i}\frac{\partial}{\partial S_j}R$ we obtain
\begin{equation}
M(z) T^{(1)}(z) +[\frac{1}{3} (T^{(0)}(z))^2] +12 H(z,z)=0
\label{t1t2}
\end{equation}
Here the
term indicated in the square bracket goes as $N^2$ (note that only the
$N_i$ dependent term in $T^{(0)}$ contribute since $G^2 w_{\alpha}$ is
trivial) and therefore represents genus 0 contribution. On the other
hand the term proportional to $H(z,z)$ does not come with any factors
of $N_i$ as is seen from the explicit solution given in
(\ref{rtsolution}), (\ref{tik}) and (\ref{cklsolution}). This contribution
therefore comes from genus 1. Writing $T^{(1)}=T^{(1)}_0 + T^{(1)}_1$
where the subscript denotes the genus, we have the following solution
to the above equation:
\begin{eqnarray}
T^{(1)}_0 (z)&=& -\frac{1}{6} N_i \frac{\partial}{\partial S_i} T^{(0)}(z)
\nonumber\\
T^{(1)}_1 (z)&=& -\frac{12}{(2 R^{(0)}(z)-V'(z))}[H(z,z) + c^{(1)}(z)]
\label{t11}
\end{eqnarray}
where $c^{(1)}(z)$ is a polynomial of degree $n-2$ and is uniquely
determined by the requirement that the contour integrals of
$T^{(1)}_1(z)$ around every branch cut vanishes. In the next
subsection we will show that this is exactly the answer the Matrix
model provides. 

Let us note that the genus 0 contribution $T^{(1)}_0$ above can be absorbed in
$T^{(0)}$ by a field redefinition 
\begin{equation}
S_i \rightarrow S_i +\frac{1}{6} G^2 N_i
\end{equation}
In particular this means that the contribution of $T^{(1)}_0$ to the
effective superpotential can be absorbed by the above field
redefinition into the original genus 0 effective superpotential in the
absence of the gravitational field. This also implies that this term
does not contribute to the superpotential when evaluated at the
classical solution of $S_i$ in agreement with the statement made in 
\cite{Ooguri:2003tt}.

\subsection{Comparison with the Matrix model results}

In the Matrix model a systematic approach to computing higher genus
contributions has been developed in \cite{Ambjorn:1993gw,Akemann:1996zr}, 
however in the
following we will rederive their results in a way parallel to the
gauge theory discussion above. This will make the comparison between
the two very transparent. Consider a hermitian matrix model with
action given by  $S=\frac{\hat{N}}{g_m}\sum_k 
\frac{g_k}{k}{\rm{tr}}M^k \equiv \frac{\hat{N}}{g_m} V$, where $M$ is a
hermitian ${\hat{N}}\times {\hat{N}}$ matrix.
In Matrix model the resolvent $\Omega(z)\equiv
\frac{g_m}{\hat{N}}\rm{tr} \frac{1}{z-M}$ 
satisfies a loop equation similar to 
gauge theory $R(z)$. 
\begin{equation}
\langle \Omega(z)\rangle ^2 - I(z)\langle \Omega(z)\rangle + \langle \Omega(z)\Omega(z)\rangle_c
=0
\end{equation} 
Here $I(z)$ is the same integral operator as in the gauge theory
discussion above.
In the large $\hat{N}$ limit, the two point function
factorizes. However in the subleading order in $1/{\hat{N}}^2$ the connected
part of the two point function (in fact the planar connected graph) 
contributes and  which in turn yields  the genus 1 
contribution to the resolvent via the above equation.
By definition 
\begin{eqnarray}
\langle \Omega(z) \Omega(w)\rangle_c &=& \frac{1}{\hat{N}^2}\sum_{k=0}^{\infty} \frac{k}{z^{k+1}}\frac
{\partial}{\partial g_k} \langle \Omega(w)\rangle 
\nonumber \\
&\equiv& \frac{1}{\hat{N}^2} {\cal O}(z)
\langle \Omega(w)\rangle
\label{c2pt}
\end{eqnarray}

Let us expand the 1-point function as :
\begin{equation}
\langle \Omega(z)\rangle = \Omega_{(0)}(z) +\frac{1}{\hat{N}^2} 
\Omega_{(1)}(z) +...
\end{equation}
where dots represent terms of higher order in
$1/{\hat{N}^2}$. Inserting these expansions in the above equations we
get:
\begin{eqnarray}
\Omega_{(0)}(z)^2 -I(z) \Omega_{(0)}(z) &=& 0 \nonumber\\
(2 \Omega_{(0)}(z) -I(z)) \Omega_{(1)}(z) &+& {\cal O}(z) \Omega_{(0)}(z)=0 
\label{r0r1eq}
\end{eqnarray}
Now we need to solve for ${\cal O}(w) \Omega_{(0)}(z)$. This can be done by
applying the
differential operator ${\cal O}(w)$ on the first equation of
(\ref{r0r1eq}). To this end we need the following identity:
\begin{equation}
{\cal O}(w) V'(y) =\sum_{k=1}^{\infty} \frac{k}{w} (\frac{y}{w})^k =
\frac{1}{(w-y)^2}
\end{equation}
which is valid for $|w| > |y|$.  It follows that 
\begin{eqnarray} 
\int_{C_z} dy[{\cal O}(w) V'(y)]\frac{\Omega_{(0)}(y)}{y-z} &=&
  \int_{C_z, |y| <|w|}
  dy \frac{1}{(w-y)^2}
\frac{\Omega_{(0)}(y)}{y-z} \nonumber\\ &=&
\partial_{w} \frac{\Omega_{(0)}(z)-\Omega_{(0)}(w)}{z-w}.
\end{eqnarray}
Using this, we obtain the following equation by applying ${\cal O}(w)$ 
on the first equation of (\ref{r0r1eq})
\begin{equation}
(2 \Omega_{(0)}(z) -I(z)) {\cal O}(w) \Omega_{(0)}(z) - \partial_w
\frac{\Omega_{(0)}(z)-\Omega_{(0)}(w)}{z-w}.
\end{equation}
Since $\Omega_{(0)}$ of the matrix model is the same as the $R^{(0)}$ for
the gauge theory, we see that  ${\cal O}(w)\Omega_{(0)}(z)$ satisfies the
same equation (\ref{rteq}) as $3 \langle R(z) T(w)
\rangle_c$.
We now impose the conditions
\begin{equation}
\int_{C_i} dz  {\cal O}(w)\Omega_{(0)}(z) = \int_{C_i} dw  {\cal
  O}(w)\Omega_{(0)}(w)=0
\end{equation}
which are the analogues of the equations (\ref{contour}). It follows
from  the discussion of uniqueness that $ {\cal O}(w)\Omega_{(0)}(z)$ is 
equal to $3 H(z,w)$. Note that as we showed in the last subsection,
$H(z,w)$
is symmetric in $z$ and $w$. This is consistent with the fact that 
 $ {\cal O}(w)\Omega_{(0)}(z)$ is symmetric in $z$ and $w$. 
Finally, substitution of   
 $ {\cal O}(z)\Omega_{(0)}(z)$ in the second equation of
(\ref{r0r1eq}), 
results in an equation for $\Omega_{(1)}(z)$ which is identical to
that for the genus 1 part of the gauge
theory
equation (\ref{t1t2}) for $T^{(1)}_1$. 
Using the fact that the integral of $\Omega_{(1)}(z)$ around every branch cut
is zero, we
conclude, from the uniqueness of the solution, that 
\begin{equation}
\Omega_{(1)}(z)=\frac{1}{4}T^{(1)}_1(z),
\label{gmequiv}
\end{equation}
with the right hand side being the genus 1 part of the solution given in (\ref{t11}).
While in the matrix model the $\frac{1}{\hat{N}^2}$ correction to the 
effective potential is obtained by integrating the asymptotic
expansion of $\Omega_{(1)}(z)$ with respect to the couplings $g_k$, the
order $G^2$ correction to the effective superpotential in gauge theory
is
obtained by integrating the asymptotic expansion of $T^{(1)}_1(z)$
with respect to the coupling constants $g_k$ (we already argued in the
last
subsection that the genus 0 contribution coming from $T^{(1)}_0$ can
be absorbed by a field redefinition of $S_i$). Eq.(\ref{gmequiv})
implies therefore that the genus 1 contribution to the effective
potential in matrix model is equal to the genus 1 contribution to the
order $G^2$ term in the gauge theory effective superpotential. 
In fact, the relative coefficient $4$ in eq.(\ref{gmequiv}) is exactly
reproduced if one follows the numerical factors in the diagrammatic
computations given in  \cite{Ooguri:2003tt}.

\section{Conclusions}
In this paper we have analyzed the problem of computing the
first non-trivial gravitational corrections to the 
effective superpotential, $\Gamma_1=\int d^4x d^2\theta
G^2 F_1(S)$,  
resulting from integrating out an
adjoint scalar superfield with polynomial tree level    
superpotential.and minimal coupling to $N=1$ supergravity.
Whereas the problem has been earlier analyzed by \cite{Ooguri:2003tt}
who showed, by using diagrammatic techniques, that
the effective superpotential is in fact given by the genus one 
correction to free energy in the corresponding matrix model,
we have here considered the issue from the point of view of
``loop equations'' arising from the generalized Konishi anomaly 
in presence of an $N=1$ gauge $W_\alpha$ and  supergravity 
 $G_{\alpha\beta\gamma}$ background superfields, whose
lowest components are the gaugino and the gravitino field strength
respectively. It should be possible to generalize these methods to the
case of
other classical groups, as well as different representations for the
chiral
fields. 

From this point of view, the appearance of non-planar
(genus one) diagrams is related to the lack of the usual factorization
property of gauge invariant correlators
in the presence of a non-trivial supergravity background,   
The fact that a non-trivial supergravity background deforms
the chiral ring was stressed and used explicitly in the 
work by \cite{Ooguri:2003tt}, 
where the failure of the factorization property   
was implicit. In this paper we have exploited both these facts and 
included also the supergravity contribution to the Konishi anomaly. 
First, we have estimated various two-and three- point connected
correlators using the modified chiral ring relations, then we have derived
loop equations which, due to the absence of factorization, involve
connected two-point functions of ``resolvents'' $R(z)$, $T(z)$ and
$w_\alpha(z)$, generalizing the equations
obtained by \cite{Cachazo:2002ry} 
in two ways: first because we found
additional contributions of order $G^2$, and 
second because we found it useful to consider
two-point functions involving operators at two different points
in the complex plane. We have then shown that these loop equations 
satisfy quite non-trivial consistency conditions and that there is a unique
solution for the non-vanishing connected two-point functions. 
Using these results, we have then
solved for the ${\cal O}(G^2)$, genus one, 
correction to the superpotential and found 
agreement with the matrix model result. We also find at ${\cal
  O}(G^2)$,
a genus zero contribution to the superpotential, which can be removed
by a field redefinition.       

Concerning the issue of
factorization (i.e. position independence) of 
chiral correlators in a supergravity
background, we have observed in section 3 that our
results do not contradict the known fact that 
this property holds in
$N=1$ gauge theories on K\"{a}hler manifolds 
(with non-trivial twisting if the manifold is not Hyper-Kahler)
\cite{Witten:1994ev},
since we indeed find that for purely bosonic background
connected correlators vanish.

Finally, it is a non-trivial open problem to
extend the above results to higher genera. Indeed,
whereas the chiral ring deformation that we have here employed 
follows directly from standard $N=1$ supergravity
tensor calculus.  
it has  been shown by \cite{Ooguri:2003qp}, using diagrammatic
techniques, that in order to capture these higher order corrections one
has to modify the chiral ring relations rather drastically,
i.e. one has to include a
self-dual two form $F_{\alpha\beta}$, which
is a remnant of the graviphoton field strength of the
parent $N=2$ supergravity. The modified relation,
$\{W_\alpha,W_\beta\}=
2G_{\alpha\beta\gamma}W^\gamma+F_{\alpha\beta}$,.
does not have a conventional interpretation in $N=1$
supergravity. It would be interesting to see how this
non-standard modification can be incorporated in the approach
followed here, based on generalized Konishi anomaly relations.      

\acknowledgments
It is a pleasure to thank George Thompson for numerous valuable 
discussions and for collaboration in the early phase of this project.
The work of the authors is supported in part by EEC contract EC
HPRN-CT-2000-00148.

\appendix

\section{Conventions}

In this paper we follow conventions of \cite{Wess:1992cp}. Raising and
lowering of spinor indices are done by the $\epsilon$ tensor as
follows
\bea{con1}
W^\alpha = \epsilon^{\alpha\beta} W_\beta, \;\;\;\;
W_\alpha = \epsilon_{\alpha\beta} W^\beta, \\ \nonumber
\epsilon^{\alpha\beta} \epsilon_{\beta\alpha} = 2, \;\;\;\;
\epsilon^{\alpha\beta} \epsilon_{\beta\alpha'} = 
\delta^{\alpha}_{\alpha'}. 
\eea
We also define the following products of the $N=1$ gauge multiplet and
the $N=1$ Weyl multiplet
\bea{con2}
W^2 &=& W^\alpha W_\alpha = - W_\alpha W^\alpha, \\ \nonumber
G^2 = G^{\alpha\beta\gamma} G_{\alpha\beta\gamma},  &\;&\;\;\;
G_{\delta\gamma\alpha} G^{\delta\gamma}_{\;\;\;\;\beta} = - G_{\delta\gamma\beta}
G^{\delta\gamma}_{\;\;\;\;\alpha} = \frac{\epsilon_{\alpha\beta}}{2} 
G^2. 
\eea

\section{The gravitational contribution to Konishi
anomaly} 

In this appendix we fix the normalization constant of the
pure gravitational contribution to the the Konishi anomaly  which was used in
\eq{Gan}. 
The Konishi anomaly equation including the pure gravitational contribution 
in superspace is given by \cite{Konishi:1988mb,Magnoli:1990qh}
\be{fulkon}
\bar{D}^2 (\bar{\Phi} e^V \Phi) = \frac{1}{32\pi^2} \rm{Tr}_{\rm{Ad}}
(W^2) + \alpha \frac{1}{32\pi^2} G^2
\ee
Here $\alpha$ is the unknown normalization constant which we will fix below.
The $\theta^2$ component of the above equation together with its
anti-holomophic counterpart should
reduce to the familiar equation of the
chiral anomaly including the gravitational contribution given below.
\be{anbos}
\partial_\mu (\bar{\psi} \bar{\sigma}^\mu \psi) - 
\partial_\mu (\psi \sigma^\mu \bar{\psi} )
= \frac{1}{32\pi^2} \frac{i}{2} \epsilon^{mnlk} 
{\rm{Tr}}_{ \rm{Ad} } ( F_{mn} F_{lk}) +
\frac{1}{32\pi^2}\frac{1}{24} \frac{i}{2} \epsilon^{mnlk} R_{mnst}
R^{st}_{\;\;\;lk}
\ee
The coefficients in the above equation have been obtained from 
\cite{Alvarez-Gaume:1985dr}
Our strategy to fix the normalization constant $\alpha$ 
will be to extract the contribution of $R\wedge R$ from the superspace
equation in \eq{fulkon} and require it to agree with the coefficient in
\eq{anbos}.
From \cite{Wess:1992cp}
the lowest component of the $N=1$ Weyl multiplet  
starts off as the gravitino field strength and is given by
\be{low}
G_{\alpha\beta\gamma} = \frac{1}{12} (
\sigma^{ab}_{\alpha\beta} \psi_{ab\gamma} + 
\sigma^{ab}_{\beta\gamma} \psi_{ab\alpha} +
\sigma^{ab}_{\gamma\alpha} \psi_{ab\beta}),
\ee
where we  have set the auxiliary field in the above formula to zero, as we are
working on shell and $a, b$ refer to the local Lorentz indices. 
The gravitino field strength is defined by 
\be{def-gravf}
\psi_{ab}^\alpha = \hat{D}_a \psi_b ^\alpha - \hat{D}_b \psi_a^\alpha, 
 \;\;\;\;\; 
\hat{D}_a \psi_b^\alpha = \del_a \psi_b^\alpha + \psi_b^\beta
\omega_{n\beta}^\alpha,
\ee
where $\omega_{n\beta}^\alpha$ is the spin connection. As we are
working with an on shell background, equations of motion for the
gravitino imply $\sigma^{ab}_{\alpha\beta}\psi_{ab\gamma} =
\sigma^{ab}_{\alpha\gamma}\psi_{ab\beta}$. Therefore on shell,
the lowest component of the Weyl multiplet is given by
\be{lowon}
G_{\alpha\beta\gamma} = \frac{1}{4} \sigma^{ab}_{\alpha\beta}
\psi_{ab\gamma}
\ee
The supersymmetric transformation on the gravitino field strength is given by
\be{susy-trans}
\delta \psi_{mn}^\alpha  = 
- \xi^\beta R_{mnab} \sigma^{ab\alpha}_\beta + \cdots
\ee
The  dots in the above equation all refer to terms that involve the fermions, 
which are not of interest for the present purpose of determining the
coefficient of $R\wedge R$. Substituting this variation in \eq{lowon}
we obtain
\be{theta-comp}
\delta G_{\alpha\beta\gamma} = \frac{1}{4} 
\sigma^{ab}_{\alpha\beta} \sigma^{cd}_{\delta\gamma} \xi^\delta R_{abcd} 
\ee
The $\theta^2$ component of 
$G^2$ contains the $R\wedge R$ term, which is imaginary and thus 
contributes to the anomaly. This is given by
\be{imag}
G^2 |_{\theta^2} =
\frac{i}{2} \frac{1}{16} \epsilon^{a'b'ab}
R_{a'b'cd} R_{ab}^{cd}
\ee
The total contribution to the anomalous current 
is obtained by subtracting this
out with the $G_{\dot{\alpha}\dot{\beta}\dot{\gamma} }$, the anti-holomorphic
contributions. 
Therefore the coefficient of the $R\wedge R$ term from $W^2$
is $1/8$. Comparing with the coefficient of $R\wedge R$ in \eq{anbos}
we see that $\alpha =1/3$ in \eq{fulkon} in order to reproduce the chiral
anomaly, including the gravitational contribution.

%

\bibliographystyle{utphys}
\bibliography{konishi}
\end{document}